\documentclass[acmsmall,review=false,screen]{acmart}

\setcopyright{none}
\renewcommand\footnotetextcopyrightpermission[1]{}
\AtBeginDocument{%
  }

\setcopyright{acmlicensed}
\copyrightyear{2018}
\acmYear{2018}
\acmDOI{XXXXXXX.XXXXXXX}

\acmJournal{JACM}
\acmVolume{37}
\acmNumber{4}
\acmArticle{111}
\acmMonth{8}
\usepackage[most]{tcolorbox}
\usepackage{xcolor}
\newcounter{mybox}

\usepackage[ruled,vlined,linesnumbered]{algorithm2e}
\SetKwFunction{FCNFLemmaSynthesizer}{2CNFLemmaSynthesizer}
\SetKwFunction{FLemmaValidator}{LemmaValidator}
\SetKwFunction{Overall}{LimICE}
\SetKwProg{Fn}{Function}{:}{}
\usepackage{multirow}
\usepackage{tabularx}
\usepackage{siunitx}
\usepackage{makecell}
\usepackage{siunitx}
\usepackage{stmaryrd}
\usepackage{hyperref}
\usepackage{minted}
\usepackage{listings}
\usepackage{float}
\newfloat{mybox}{htbp}{lob}

\usepackage{threeparttable}

\begin{document}

%%
%% The "title" command has an optional parameter,
%% allowing the author to define a "short title" to be used in page headers.
% \title{LimICE: Integrating LLM into ICE Framework for Efficient Loop Invariant Inference}
\title{LimICE: Integrating LLM into ICE Framework for Efficient
Loop Invariant Inference}
%%
%% The "author" command and its associated commands are used to define
%% the authors and their affiliations.Do Not Use This Code
%% Of note is the shared affiliation of the first two authors, and the
%% "authornote" and "authornotemark" commands
%% used to denote shared contribution to the research.
\author{Kai Fan}
%\authornote{Both authors contributed equally to this research.}
\email{fankai20@nudt.edu.cn}
%\orcid{1234-5678-9012}
\affiliation{%
	\institution{National University of Defense Technology}
	\country{China}
}
\author{ShiWen Yu}
%\authornotemark[1]
\email{yushiwen14@nudt.edu.cn}
\affiliation{%
	\institution{National University of Defense Technology}
	\country{China}
}

\author{Guangsheng Fan}
\email{guangshengfan@nudt.edu.cn}
\affiliation{%
	\institution{National University of Defense Technology}
	\country{China}
}

\author{HaoAng Chi}
\email{haoangchi618@gmail.com}
\affiliation{%
	\institution{National University of Defense Technology}
	\country{China}
}

\author{WanWei Liu}
\email{wwliu@nudt.edu.cn}
\affiliation{%
	\institution{National University of Defense Technology}
	\country{China}
}

\author{Ji Wang}
\email{wj@nudt.edu.cn}
\affiliation{%
	\institution{National University of Defense Technology}
	\country{China}
}

%%
%% By default, the full list of authors will be used in the page
%% headers. Often, this list is too long, and will overlap
%% other information printed in the page headers. This command allows
%% the author to define a more concise list
%% of authors' names for this purpose.
\renewcommand{\shortauthors}{Trovato et al.}

%%
%% The abstract is a short summary of the work to be presented in the
%% article.
\begin{abstract}
  Loop invariant synthesis is a fundamental problem in program verification, yet the inherent undecidability makes it highly challenging. Recent studies have increasingly employed various machine learning techniques to generate loop invariants. However, most of these methods adopt a monolithic approach. Due to the inability to strictly constrain the learning process, learning-based methods struggle to simultaneously consider all necessary conditions and generate complete invariants when tackling complex problems. In fact, a loop invariant is often an ordered sequence of lemmas, rather than a single invariant formula. This motivates us to propose \textbf{Incremental ICE}, a novel learning framework for incremental synthesis. Our framework integrates the incremental philosophy of IC3 into the general invariant learning framework ICE. By defining a lemma-specific learning objective and introducing a counterexample filtering mechanism, we can achieve sound incremental learning. Under this framework, we instantiate a loop invariant synthesis tool, LimICE, which leverages LLMs to generate the ordered sequence of lemmas and incorporates ICE-DT as a fallback mechanism to complement the lemma sequence. Experiments on 367 linear benchmarks and 50 nonlinear benchmarks demonstrate the effectiveness of the proposed approach. LimICE solves 349 (out of 367) linear problems on an average of 15.2 seconds and 47 (out of 50) nonlinear problems on an average of 8.8 seconds. Compared to the state-of-the-art LLM-based baseline, our approach solves 12-24$\%$ more instances while running 36-63$\%$ faster across linear and nonlinear benchmarks. LimICE also consistently outperforms strong non-LLM baselines and solves at least 86 and 27 additional instances on the linear and nonlinear benchmarks, respectively. In addition, we conduct experiments on different LLM backends, and LimICE can still solve 326 linear problems and 32 nonlinear problems on a 7B model, demonstrating the stability of our method.
  % Recent studies have demonstrated the great potential of LLMs in assisting loop invariant generation. A promising way is to query LLMs generate atomic predicate literals and then combine them into full invariants in a counterexample-driven manner. However, this paradigm is hindered by two shortcomings: (1) it relies solely on LLM to generate all the predicates that make up the invariants, and missing predicates can fundamentally limit the approach; (2) when invariant combination fails, it is hard to locate the failure reason and provide the LLM with structured feedback. In this paper, we propose an incremental approach to address these issues. The goal of the incremental approach is to continuously infer relatively inductive lemmas rather than to obtain the monolithic invariant. In our work, the inference of each lemma is relatively simple and can have different sources. Meanwhile, inferred lemmas can provide structured guidance for LLMs. To implement the incremental approach, we introduce a counterexample filtering mechanism and present an incremental learning framework, Incremental ICE. Under this framework, we instantiate an effective loop invariant inference method, LimICE, which simultaneously utilizes the LLM and ICE-DT to infer different lemmas. 
\end{abstract}

%%
%% The code below is generated by the tool at http://dl.acm.org/ccs.cfm.
%% Please copy and paste the code instead of the example below.
%%
\begin{CCSXML}
<ccs2012>
 <concept>
  <concept_id>00000000.0000000.0000000</concept_id>
  <concept_desc>Do Not Use This Code, Generate the Correct Terms for Your Paper</concept_desc>
  <concept_significance>500</concept_significance>
 </concept>
 <concept>
  <concept_id>00000000.00000000.00000000</concept_id>
  <concept_desc>Do Not Use This Code, Generate the Correct Terms for Your Paper</concept_desc>
  <concept_significance>300</concept_significance>
 </concept>
 <concept>
  <concept_id>00000000.00000000.00000000</concept_id>
  <concept_desc>Do Not Use This Code, Generate the Correct Terms for Your Paper</concept_desc>
  <concept_significance>100</concept_significance>
 </concept>
 <concept>
  <concept_id>00000000.00000000.00000000</concept_id>
  <concept_desc>Do Not Use This Code, Generate the Correct Terms for Your Paper</concept_desc>
  <concept_significance>100</concept_significance>
 </concept>
</ccs2012>
\end{CCSXML}

\ccsdesc[500]{Software and its engineering~Formal software verification}
% \ccsdesc[300]{Do Not Use This Code~Generate the Correct Terms for Your Paper}
% \ccsdesc{Do Not Use This Code~Generate the Correct Terms for Your Paper}
% \ccsdesc[100]{Do Not Use This Code~Generate the Correct Terms for Your Paper}

%%
%% Keywords. The author(s) should pick words that accurately describe
%% the work being presented. Separate the keywords with commas.
\keywords{Program verification, loop invariant, incremental approach, LLM, ICE framework}

% \received{20 February 2007}
% \received[revised]{12 March 2009}
% \received[accepted]{5 June 2009}

%%
%% This command processes the author and affiliation and title
%% information and builds the first part of the formatted document.
\maketitle

\section{Introduction}

Program verification~\cite{induction,hoarelogic} is an important technique for ensuring the correctness of software behavior, especially in security-critical areas. A key objective is to prove that a program satisfies its specification for all possible executions, rather than for a limited set of test inputs. During this process, loop invariants play a central role. In a nutshell, a loop invariant is a specification that holds before and after each iteration of the loop. In deductive verification, synthesizing a proper invariant implies that the problem is solved. However, synthesizing loop invariants is intractable, and it cannot be automatically done in general.

A kind of classic method adopts the “guess-and-check” approach~\cite{code2inv,cln2inv,gcln,lipus}. These methods iteratively generate candidate loop invariants and check the correctness using an SMT solver. The ICE framework~\cite{ice,icedt,hornice,icearray} is a representative of this approach. It treats the source code as a black box and considers only the counterexamples returned by the SMT solver at each iteration, which partially characterize the correct loop invariant. However, when solving complex problems, excessive accumulation of counterexamples can over-constrain the synthesis process, leading to degraded scalability, loss of generality, and even synthesis failure.

% the template-based ICE framework~\cite{ice,icedt,hornice,icearray,lipus} is an important technical route. This type of method applies a \textit{guess-and-check} approach, which maintains a set of counterexamples (CE set) and iteratively proposes a candidate loop invariant that can pass all counterexamples. If the candidate can be proven to be true, then we get a correct loop invariant; otherwise, the prover will return new counterexamples, then we add them into the CE set and repeat the previous procedure.

Recently, synthesizing loop invariants using LLMs~\cite{loopy,canllm,llm4mc,lam4inv,clause2inv,lemur,llmbenchmarks} has received widespread attention. However, LLMs often struggle to generate completely correct loop invariants directly. Therefore, existing methods intend to guide LLMs in repairing loop invariants or to obtain valid invariants by filtering or combining the outputs of LLMs. An outstanding implementation is the \textit{generate-combine-check} framework implemented by Cao et al~\cite{clause2inv}, which divides the loop invariant synthesis task into sub-tasks of literal generation and logical combination. This framework allows LLM to focus on generating potentially useful literals and uses a search method to solve the logical connection problem. The drawback lies in that it cannot provide sufficient guidance for LLMs, which causes the literal generation process to be decoupled from the entire loop invariant synthesis. Moreover, as the number of literals grows, it becomes increasingly difficult to combine a complete loop invariant in a single attempt.

% relies solely on LLM to generate literals and will fail when the LLM cannot generate new literals.

\textit{Our work.} In this paper, we propose an incremental learning framework to address the aforementioned challenges. A complete loop invariant is typically not a single formula, but an ordered sequence of lemmas, i.e., $\psi_1(\bar{x}),\dots,\psi_n(\bar{x})$. These lemmas can be categorized into bounding lemmas, which constrain variable ranges, and essential lemmas, which capture semantic information~\cite{loopanalysis}. Within the sequence, lemmas are often interdependent: partial essential lemmas rely on bounding lemmas to maintain their inductiveness—a relationship defined in IC3~\cite{ic31,ic32,ic33} as relative inductiveness. Deriving the lemma sequence incrementally enhances both the convergence of the synthesis process and the reusability of the results. Even if we do not obtain sufficient lemmas to prove the postcondition, the acquired lemmas can still serve as specifications for the loop header. To enable learning-based incremental synthesis, we design a lemma-specific learning objective and a counterexample filtering mechanism, which extend the general invariant learning framework ICE to Incremental ICE. When a new lemma is derived, it is used to filter relevant inductive counterexamples and negative counterexamples, thereby guiding the learning process to follow the lemma sequence while preventing excessive expansion of the counterexample set.

Under the Incremental ICE framework, we instantiate an efficient invariant synthesis tool called LimICE, which leverages LLMs to generate lemmas and integrates ICE-DT~\cite{icedt} as a fallback mechanism. For the application of LLMs, we adopt the generate-combine-check paradigm~\cite{clause2inv}. Specifically, the LLM first generates atomic literals, which are subsequently combined into candidate lemmas. Based on the proposed lemma-specific learning objective, we design a lemma search algorithm operating in the CNF hypothesis space, which enables efficient and non-redundant exploration. Due to the ability of the incremental approach to progressively narrow the state space, our search algorithm exhibits robust performance on complex problems, especially on nonlinear problems. Furthermore, the lemma sequence can indicate the proof progress, thus providing more guidance for LLMs, which reconnects the literal generation process with the entire invariant synthesis process. Additionally, benefiting from the modular nature of the incremental approach, when LLM-based synthesis fails, we can call ICE-DT to synthesize the missing lemma from the retained counterexamples. This dual strategy enables our approach to combine the generalized code understanding capability of LLMs with the precise synthesis capability of ICE-DT.

\textit{Evaluation.} To thoroughly evaluate LimICE, we have conducted experiments on 367 linear benchmarks and 50 nonlinear benchmarks. The linear benchmarks consist of 313 instances from LaM4Inv~\cite{lam4inv} and 54 new instances from the CHC-Comp~\cite{chccomp}. The nonlinear benchmarks are from Clause2Inv~\cite{clause2inv}. We compare the performance with eight state-of-the-art (SOTA) methods, including LoopInvGen~\cite{loopinvgen}, ICE-DT~\cite{icedt}, ICE-DT-Interval~\cite{icedtinterval}, Code2Inv~\cite{code2inv}, LIPuS~\cite{lipus}, CLN2INV~\cite{cln2inv}, LaM4Inv~\cite{lam4inv}, and Clause2Inv~\cite{clause2inv}. The experiments show that LimICE outperforms other methods both in linear benchmarks and nonlinear benchmarks. It solves 349 linear benchmarks on an average of 15.2 seconds and 47 nonlinear benchmarks on an average of 8.8 seconds. Meanwhile, we conduct comparative experiments across all LLM-based methods on different LLM backends. The results show that LimICE expresses stable performance across different models. Even on the 7B model of QWen, we can still solve 326 linear benchmarks on an average of 34.4 seconds and 32 nonlinear benchmarks on an average of 45.6 seconds. In addition, we design controlled experiments that constrain LLM outputs to validate the effectiveness of the incremental combination strategy, along with a series of ablation studies to assess the contribution of each component of our approach. The details are shown in Section~\ref{sec: evaluation}.

Our contributions can be summarized as follows.

\begin{itemize}
    \item We incorporate the incremental philosophy of IC3 and extend the general invariant learning framework ICE to Incremental ICE, which is a sound framework for incremental invariant learning. This framework is orthogonal to existing methods.
    \item We instantiate an efficient invariant synthesis tool called LimICE, which achieves SOTA performance even under a standalone incremental combination strategy. Building upon this, we introduce a warm start mechanism and an ICE-DT fallback mechanism to further improve efficiency and effectiveness.
    \item We curate new benchmarks for loop invariant synthesis and conduct experiments on them. The results confirm that LimICE solves the most benchmarks with competitive efficiency compared to the SOTA baselines.
\end{itemize}

%that integrates the LLM-based generate-combine-check method and the ICE-DT method. Meanwhile, we design a novel feedback mechanism and a warm start mechanism, which enable more effective and efficient loop invariant inference.

The rest of the paper is organized as follows. Section~\ref{sec: preliminary} introduces the background and related technologies. Section~\ref{sec: incremental ice framework} introduces the Incremental ICE framework. Section~\ref{sec: the proposed approach} describes the proposed approach. Section~\ref{sec: evaluation} demonstrates the evaluation results and conducts case studies. Section~\ref{sec: discussion} discusses threats to validity. Section~\ref{sec: related work} introduces the related work. Section~\ref{sec: conclusion} concludes the paper.

\section{Preliminary}\label{sec: preliminary}

\subsection{Loop Invariant}\label{sec:Loop Invariant}

The purpose of synthesizing loop invariants is to make sure that the program's behavior satisfies the expected property. Let $P$ and $Q$ denote the predicates on the program variables, and let $S$ denote this program. Then we can describe the property by a \textit{Hoare triple} $\{P\}  S  \{Q\}$~\cite{hoarelogic}. For a program in the form $\texttt{while} \ B \ \texttt{do} \ S$, a proper loop invariant $I$ that can valid the \textit{Hoare triple} must satisfy:

\begin{equation}
    \frac{P \Longrightarrow I \ (pre) \quad\{I \wedge B\} S\{I\} \ (indu) \quad(I \wedge \neg B) \Longrightarrow Q \ (post)}{\{P\} \texttt{ while } B \texttt{ do } S \ \{Q\}}.
    \label{equ:invariant}
\end{equation}
Here, $P$ is the \textit{precondition}, Q is the \textit{postcondition}, $S$ is the loop body, and $B$ is the loop condition. The above rule can be divided into three parts:

\begin{itemize}
    \item \textit{pre.} The loop invariant must hold when entering the loop. This means that the invariant is true when entering the loop from any state that satisfies the \textit{precondition};
    \item \textit{indu.} If the loop invariant holds before an iteration of the loop, it will still hold after executing the loop body. This means that the invariant remains true throughout the entire loop execution;
    \item \textit{post.} When the loop terminates, the invariant can prove that the \textit{postcondition} holds. This means that the invariant is strong enough for the proving problem.
\end{itemize}

Based on the above three premises, we can prove that whenever the program $S$ starts from a state satisfying $P$, $Q$ will always hold when the loop ends. In practice, the correctness of these conditions can be examined by SMT solvers.

\subsection{ICE Framework}\label{sec:ICE Framework}

The ICE framework~\cite{ice,icedt,hornice,lipus,icearray} is a learning-based approach to finding invariants. This framework consists of two components. One acts as a \textit{teacher} who is responsible for verifying the correctness of the candidate invariant and generating counterexamples. The other acts as a \textit{learner} who learns from the counterexamples given by the teacher and proposes a candidate invariant. In each iteration, the teacher checks the correctness of three conditions in equation~\eqref{equ:invariant}. If an error occurs, the concrete value assignment for the program variables will be returned as new counterexamples. Depending on the positions where errors occur, counterexamples can be categorized into three types.

\begin{itemize}
    \item \textit{positive counterexample $(p)$. } This kind of counterexample falsifies $P \Longrightarrow I \ (pre)$. The next invariant should be true under the assignment of this counterexample;
    \item \textit{negative counterexample $(n)$. } This kind of counterexample falsifies $(I \wedge \neg B) \Longrightarrow Q \ (post)$. The next invariant should be false under the assignment of this counterexample;
    \item \textit{inductive counterexample $(i_1,i_2)$. } This kind of counterexample falsifies $\{I \wedge B\} S\{I\} \ (indu)$. The next invariant should be false under the first assignment of this counterexample or true under both assignments of this counterexample.
\end{itemize}

All counterexamples will be collected and maintained in three sets ($CE_p$,$CE_n$, and $CE_i$), and the three sets can partially characterize the property of correct loop invariants.

\subsection{Generate-Combine-Check Framework}

The \textit{Generate-Combine-Check} framework is motivated by the fact that LLMs are struggling to guess the complete invariant, while generally all the components of a correct loop invariant may appear in multiple guessing processes~\cite{clause2inv}. Unlike the guess-and-check approach, this framework first queries LLMs or other loop invariant generation tools to generate atomic literals without logical connections and stores all historically generated literals in a set. Then it uses a symbolic combiner to search for a candidate loop invariant from the combinatorial space of literals, which is then checked by an SMT solver. If the symbolic combiner cannot find new candidates, the framework returns to the generate stage to obtain new literals.

% Thus, this framework divides the loop invariant inference task into literal generation and logical combination. In literal generation, LLMs are required to generate literals without logical connectives. In literal combination, all literals generated by LLM in history are combined with logical connectives $\land$ and $\lor$.
% To guide the combination process, Clause2Inv, an implementation of this framework, uses a counterexample-driven approach. It maintains a set of counterexamples as shown in section \ref{sec:ICE Framework} and a candidate invariant set \textit{exprList}. Each time LLM generates new literals, Clause2Inv first appends them to \textit{exprList} and then combines each new literal $c_{new}$ and each old predicate $c_{old}$ as $c_{new}\land c_{old}$ and $c_{new}\lor c_{old}$, which are also appended to \textit{exprList}. After that, it will score each candidate invariant in \textit{exprList} according to the number of counterexamples they can pass and check the candidate with the highest score using an SMT solver. If the candidate is true, then a correct loop invariant is found. Otherwise, Clause2Inv refines the candidate with relatively high scores. This progress continues until there are no candidates that can pass all counterexamples.

\section{Incremental ICE Framework}\label{sec: incremental ice framework}

In this section, we introduce the incremental learning framework, \textbf{Incremental ICE}, which is a variant of the standard ICE framework. We first introduce the mathematical preliminaries of the incremental approach, followed by the technical details for implementation.

\subsection{Incremental Synthesis of IC3}

Induction is fundamental to the verification of safety properties~\cite{ic33,hoarelogic,induction}, comprising initiation and consecution, which correspond to $pre$ and $indu$ conditions of a loop invariant requirement, respectively. Consider the property described in Section~\ref{sec:Loop Invariant} —— $\{P\} \texttt{ while } B \texttt{ do } S \ \{Q\}$. To facilitate the following description, we first transform the program into a finite state system $\langle\bar{x},P(\bar{x}),T(\bar{x},\bar{x}')\rangle$. Here, $\bar{x}$ is the vector of program variables and $T(\bar{x},\bar{x}')$ describes the transition relation $\bar{x}'=\llbracket S\rrbracket(\bar{x})$. In this setting, the postcondition $Q$ to be proved in the original program is reformulated as a property $Q(\bar{x}) \lor B(\bar{x})$ that must hold throughout the executions of the finite state system. One typical tactic to prove such a property is given below.

\begin{itemize}
    \item \textit{Initiation:} prove that the property holds initially: $P(\bar{x})\Rightarrow  Q(\bar{x}) \lor B(\bar{x})$;
    \item \textit{Consecution:} prove that if the property holds before a transition, so does it after the transition: $(Q(\bar{x}) \lor B(\bar{x}))  \land  T(\bar{x},\bar{x}') \Rightarrow (Q(\bar{x}') \lor B(\bar{x}')) $.
\end{itemize}

If we can prove the above two conditions, there is no need to synthesize any extra loop invariant. If this is not the case, usually two strategies can be applied to deal with it~\cite{increasebook,ic33}: (1) conduct a stronger assertion $\Psi$, and (2) conduct an incremental lemma sequence $\psi_1,\dots,\psi_n$. The former is a monolithic approach; it tries to synthesize a strong loop invariant $\Psi$ in one proposal, which satisfies the following equation.

\begin{equation}
\left\{\begin{matrix}
\begin{aligned}
&P(\bar{x} ) \Rightarrow \Psi(\bar{x})
 \\
&\Psi(\bar{x}) \land T(\bar{x},\bar{x}')\Rightarrow \Psi(\bar{x}' )
 \\
&\Psi(\bar{x} )\Rightarrow Q(\bar{x}) \lor B(\bar{x})
\end{aligned}
\end{matrix}\right..
\end{equation}

This approach is often adopted by traditional automated symbolic verification techniques, whereas the incremental approach is more common in manual inference by human experts~\cite{ic33}, since it offers the advantages of reusability, better convergence, and modularity. In detail, it intends to synthesize a lemma sequence $\psi_1(\bar{x}),\dots,\psi_n(\bar{x})$, and its definition is shown below.

\begin{definition}[lemma sequence]\label{defi: lemma sequence}
A lemma sequence is an ordered sequence of assertions, where the assertions satisfy the following requirements:
\begin{itemize}
    \item each assertion holds initially, namely, for each $j$, $P(\bar{x})\Rightarrow \psi_j(\bar{x})$;
    \item each assertion satisfies the relative inductiveness: it obeys consecution when the earlier assertions are valid, that is, for each $j$,
    \begin{equation*}
        \bigwedge_{1 \leq k < j} \psi_{k}(\bar{x})\wedge \psi_j(\bar{x}) \wedge T\left(\bar{x}, \bar{x}^{\prime}\right) \Rightarrow \psi_{j}\left(\bar{x}^{\prime}\right);
        \label{equ: incremental inductive}
    \end{equation*}
    \item all the assertions together can prove the property, namely,
    \begin{equation*}
        \bigwedge_{1 \leq j \leq n} \psi_{j}(\bar{x}) \Rightarrow Q(\bar{x}) \lor B(\bar{x}).
    \end{equation*}
\end{itemize}
The final loop invariant is the conjunction of the lemma sequence, i.e., $\bigwedge_{1 \leq j \leq n} \psi_{j}$.
\end{definition}

\subsection{Incremental ICE}\label{section: Incremental ICE}

\begin{figure}[t]
  \centering
  \includegraphics[width=\linewidth]{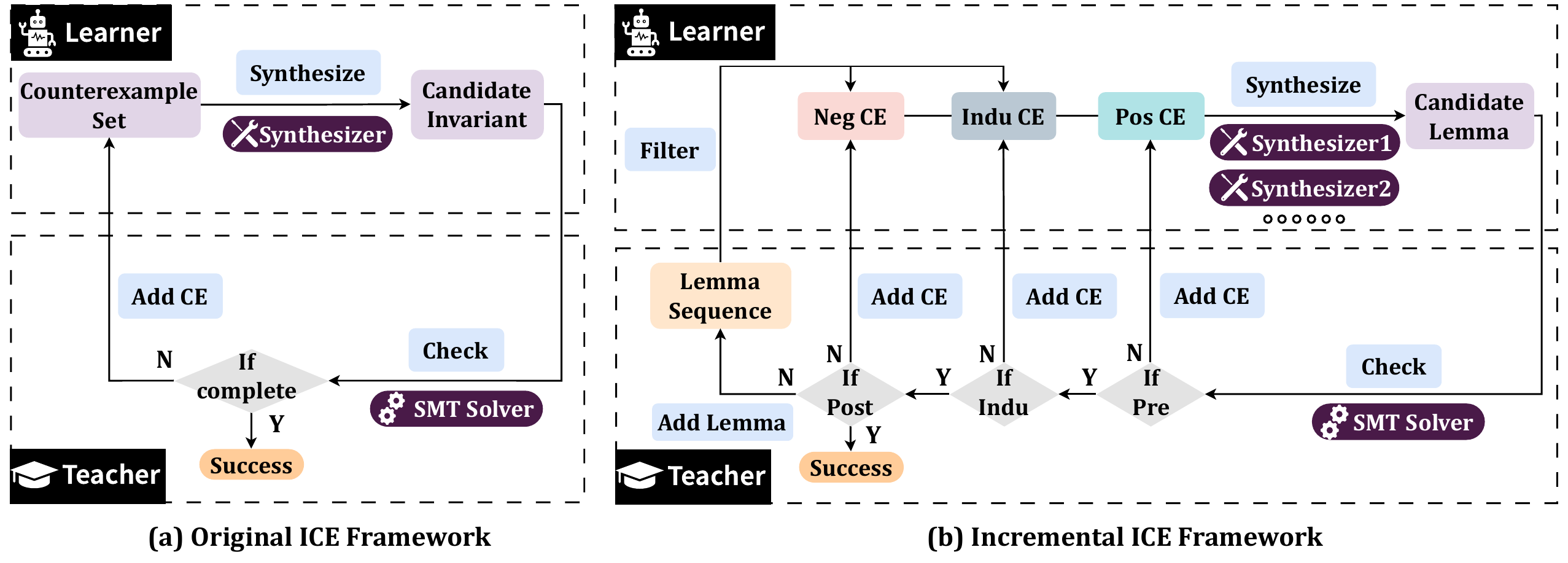}
  \caption{Comparison between the Incremental ICE framework and the Original ICE framework.}
  \label{fig: Incremental ICE}
\end{figure}

To implement an incremental approach in the ICE framework, we introduce a lemma-specific learning objective and a counterexample filtering mechanism to the original ICE framework. Figure~\ref{fig: Incremental ICE} depicts the difference between these two frameworks. The goal of the learner in incremental ICE is to propose a new candidate lemma instead of a monolithic loop invariant. Thus, the new lemma $\psi_k$ only needs to pass all positive counterexamples, all inductive counterexamples, and exclude any one of the negative counterexamples, as shown below. 

\begin{equation}
\begin{array}{ll}
\forall p \in C E_{p}, & \ \bar{x} =p \Rightarrow \psi_k(\bar{x} ) \\
\exists n \in C E_{n}, & \ \bar{x}=n \Rightarrow \neg \psi_k(\bar{x}) \\
\forall (i_1,i_2) \in C E_{i}, & \ \left(\bar{x}_1=i_{1} \wedge \bar{x}_{2}=i_{2}\right) \Rightarrow\left(\psi_k\left(\bar{x}_{1}\right) \Rightarrow \psi_k\left(\bar{x}_{2}\right)\right)
\end{array}.  \label{equ: new lemma condition}
\end{equation}
The notion of counterexamples used here follows the definition in Section~\ref{sec:ICE Framework}. Due to its modular nature, the learner can propose different lemmas using different methods.

The teacher checks whether the candidate lemma can be added to the lemma sequence according to the conditions in definition~\ref{defi: lemma sequence}. First, it checks whether the new lemma holds initially, since all lemmas in the lemma sequence should satisfy this condition. Second, it checks whether the new lemma is inductive relative to the current lemma sequence. Each lemma can shrink the state space and relax induction requirements, making it easier for the new candidate lemmas to satisfy relative inductiveness. If either of the two steps above fails, the teacher stops checking and immediately returns new counterexamples to the learner. If the candidate lemma passes the first two checks, it becomes a new lemma and will be added to the lemma sequence. When the accumulated lemmas are sufficient to prove the property, then $\bigwedge_{1 \leq j \leq k} \psi_{j}$ will be returned as a loop invariant. Otherwise, the learner will use the new lemma to filter out relevant inductive counterexamples and negative counterexamples. The specific filtering scope is defined below.

\begin{equation}
\{(i_1,i_2)\in CE_i \mid \bar x = i_1 \Rightarrow \neg \psi _k(\bar x)\} \label{equ: filter content} \ \cup \ \{n\in CE_n \mid \bar x = n \Rightarrow \neg \psi_k(\bar x)\}.
\end{equation}

We filter out these counterexamples because they cannot restrict the new lemma. The general goal of incremental ICE is to let the whole lemma sequence pass all counterexamples, as shown below.

\begin{equation}
\begin{array}{ll}
\forall p \in C E_{p}, & \ \bar{x} =p \Rightarrow \bigwedge_{1 \leq j \leq k}\psi_j(\bar{x} ) \\
\forall n \in C E_{n}, & \ \bar{x}=n \Rightarrow \neg \bigwedge_{1 \leq j \leq k}\psi_j(\bar{x}) \\
\forall (i_1,i_2)  \in C E_{i}, & \ \left(\bar{x}_1=i_{1} \wedge \bar{x}_{2}=i_{2}\right) \Rightarrow\left(\bigwedge_{1 \leq j \leq k}\psi_j\left(\bar{x}_{1}\right) \Rightarrow \bigwedge_{1 \leq j \leq k}\psi_j\left(\bar{x}_{2}\right)\right)
\end{array}. \label{equ: incremental ice goal}
\end{equation}

Since each lemma should hold initially, they should be true for all positive counterexample assignments. Thus, we will not filter out any positive counterexample. For inductive counterexample $(i_1,i_2)$, if $\psi_k$ is false for the assignment of $i_1$, then $\bar{x}=i_i\Rightarrow\neg (\psi_k(\bar{x})\land \psi'(\bar{x}))$ holds for any predicate $\psi'$. Based on this, we can further conclude that the following statement holds for any predicate $\psi'$.

\begin{equation}
    (\bar{x}_1=i_i\land \bar{x}_2=i_2)\Rightarrow (\psi_k(\bar{x}_1)\land \psi'(\bar{x}_1)\Rightarrow\psi_k(\bar{x}_2)\land \psi'(\bar{x}_2)).
\end{equation}

Thus, we can safely filter out $(i_1,i_2)$ without losing information. The same principle applies to filtering negative counterexamples. For negative counterexample $n$, if $\psi_k$ is false for the assignment of $n$, then $\bar{x}=n\Rightarrow \neg(\psi_k(\bar{x})\wedge\psi'(\bar{x}))$ holds for any predicate $\psi'$. Therefore, we can also safely filter out $n$. When the learner synthesizes new lemmas, it only needs to consider the constraints of the retained counterexamples.

\section{The Proposed Approach}\label{sec: the proposed approach}

\begin{figure}[t]
  \centering
  \includegraphics[width=\linewidth]{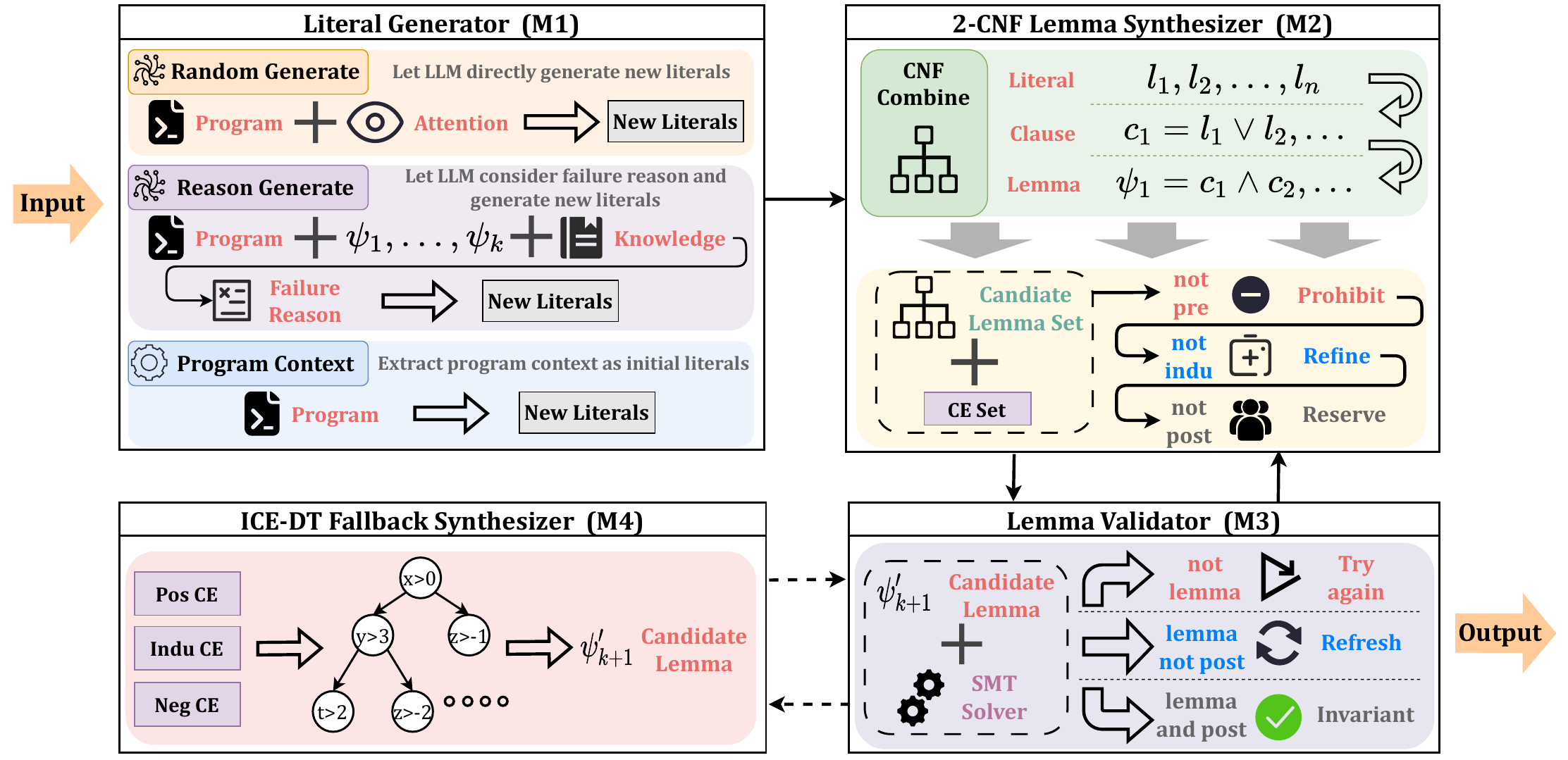}
  \caption{A Framework of LimICE.}
  \label{fig: LimICE framework}
  \vspace{-0.3em}
\end{figure}

In this section, we present LimICE, an incremental loop invariant synthesis approach under the Incremental ICE framework, which takes LLM and ICE-DT~\cite{icedt} as two different lemma synthesizers. LimICE employs the LLM-based synthesizer as the primary synthesizer and ICE-DT as the fallback synthesizer. Figure~\ref{fig: LimICE framework} presents a framework for our approach, which consists of four interacting modules: Literal Generator (M1), 2-CNF Lemma Synthesizer (M2), Lemma Validator (M3), and ICE-DT Fallback Synthesizer (M4). The first three modules form the workflow of the LLM-based learning process, while Modules M3 and M4 together form the fallback learning process. The remainder of this section first presents an overview of LimICE, followed by the details of its core components.

% \begin{enumerate}
%     \item Initially, we generate new literals in three different approaches. Before starting inference, some program contexts are extracted as initial literals for a warm start. During the inference process, we let the LLM generate new literals in a random approach and a reasoning approach.
%     \item Then, we search and check for lemmas in the combinatorial space of literals using the grammatical form of conjunctive normal form(CNF). First, construct clauses by taking pairwise disjunctions of literals and using counterexamples to filter all combinations of legal clauses. Then the shortest one reserved will be checked by the SMT solver.
%     \item If we cannot infer enough lemmas to prove the property, the ICE-DT algorithms, as a backup mechanism, will be called to synthesize the lacking lemma. And if this step also fails, the new counterexamples will be added to the set, and we will go back to step 1.
% \end{enumerate}

% The above process is repeated until the lemma sequence is powerful enough to prove the property or the computing resources are exhausted. In the sequel, we present the details of three steps.

\subsection{The Overall Algorithm}

The overall algorithm of LimICE is summarized in Algorithm~\ref{alg: LimICE}. Given the program, LimICE first initializes the data structures and performs a warm start on the \texttt{clauseSet} (line 2). Specifically, we maintain four data structures throughout the synthesis lifecycle: \texttt{clauseSet} storing historical clauses, \texttt{reserveSet} storing candidate lemmas that may be useful in the future, \texttt{lemmaSeq} storing the lemma sequence, and \texttt{CESet} storing retained counterexamples. During each iteration, LimICE first performs the LLM-based learning process. This process begins by querying the LLM to generate new literals in the random generation method (line 6). If all the new literals have already appeared in the \texttt{clauseSet}, we turn to querying the LLM in the reason generation method (line 8). Then, LimICE updates the \texttt{clauseSet} with new literals and searches for candidate lemmas that satisfy equation~\eqref{equ: new lemma condition} from the conjunction of clauses in the \texttt{clauseSet} (line 9). If the search process can propose a candidate lemma, LimICE will validate it with respect to the definition~\ref{defi: lemma sequence} of the incremental approach (line 11). If the validation succeeds, a complete invariant is returned (line 13); otherwise, the validation results are fed back into the 2-CNF Lemma Synthesizer to guide the subsequent search (line 14). The above search and validation process is iterated until the problem is solved or no new candidate lemma can be proposed, at which point LimICE switches to the fallback learning process (lines 16-18). The ICE-DT will be called to synthesize the missing lemma based on the retained counterexamples.

\subsection{LLM-Based Learning Process}

\subsubsection{Literal Generator}\label{section: Generate New Literals}

We design three ways to generate new literals. The first way is called random generation. At the start of each iteration, the LLM is prompted directly generate new literals that do not contain logic connectives. Here, we follow the approach of~\cite{clause2inv} and only modify the prompt by adding explicit constraints to the output format/content. This approach allows us to obtain logical assertions about the basic facts of the program, but to some extent separates the LLM from the overall invariant synthesis process. In the absence of effective guidance, the LLM tends to repeatedly produce literals confined to a fixed subset. This may not be enough to solve some complex problems, and thus, we design a supplementary way to generate new literals when random generation fails.

\begin{algorithm}[t]
\caption{Procedure \textsc{LimICE}}
\label{alg: LimICE}
\KwIn{Program information $\mathit{program}$, the number of ICE-DT iterations in each round $\mathit{n}$.}
\KwOut{Loop invariant $\mathit{inv}$}

\Fn{\Overall{$\mathit{program}$}}{
    $\mathit{clauseSet}\gets WarmStart(program)$\;
    $\mathit{reserveSet},\mathit{LemmaSeq},\mathit{CESet}\gets \emptyset$\;
    \While{\textbf{True}}{
        \texttt{/* }\textbf{LLM-based Learning Process}\texttt{ */}\;
        $\mathit{newLiterals}\gets RandomGenerate(program)$\;
        \If{$\mathit{newLiterals} = \emptyset$}{
            $\mathit{newLiterals}\gets ReasonGenerate(program,LemmaSeq)$\;
        }
        $\mathit{candidate}\gets 2CNFLemmaSynthesizer(\mathit{newLiterals}, clauseSet, CESet, reserveSet)$\;
        \While{$\mathit{candidate} \not = \emptyset$}{
            $\mathit{status} = LemmaValidator(\mathit{candidate},LemmaSeq,clauseSet, CESet, reserveSet)$\;
            \If{$\mathit{status} = \textbf{True} $}{
                \Return $Conjunction(LemmaSeq)$\;
            }
            $\mathit{candidate}\gets 2CNFLemmaSynthesizer(\emptyset, clauseSet, CESet, reserveSet)$\; 
        }
        \texttt{/* }\textbf{Fallback Learning Process}\texttt{ */}\;
            $status\gets ICEDTLearn(LemmaSeq,CESet,n)\qquad$\tcp{Iterate $n$ times}
            \If{$\mathit{status} = \textbf{True} $}{
                \Return $Conjunction(LemmaSeq)$\;
            }          
    }
}
\end{algorithm}

The supplementary way is called reason generation. Since the lemma sequence can indicate the current loop invariant synthesis progress, we can leverage it to provide structural guidance for LLMs. We first integrate the knowledge about the incremental approach, the lemma sequence, and the program to be verified into a prompt, and query the LLM to explain why the current lemma sequence cannot prove the property. The simplified prompt template is shown in Prompt~\ref{prompt: reason failure}. Then we prompt the LLM to generate new literals according to the failure reason. The literal generation process, which combines the two approaches, is closer to how human experts solve problems: first, attempt to prove the property based on the basic facts of the program. When this step fails, analyze the failure and identify the missing facts.

In addition, we design a warm start mechanism. Before starting synthesis, the program's pre-conditions, post-conditions, branch conditions, and inverse of branch conditions are extracted. These literals and their pairwise disjunctions will be the initial \texttt{clauseSet}. Empirical evidence suggests that these predicates significantly help to prove the property.

\begin{mybox}[t]
	\begin{tcolorbox}[title = {Prompt 1: Reason Failure Prompt}]
		\refstepcounter{mybox}
		$\#\#\#$ Your Task $\#\#\#$
		
		Your task is to explain why we cannot prove the postconditions using the given lemma sequence.
		\\
		\\
		$\#\#\#$ Task Description $\#\#\#$
		
		In program property verification, handling loops is one of the most critical steps. We aim to adopt an incremental approach to address the verification of program properties involving loops. This method computes a sequence of assertions that satisfy the following conditions: "....." 
		
		Currently, we have obtained the following sequence of assertions:
		\{Lemma Sequence\} 
		
		This sequence has been proven to satisfy the first two conditions but fails to satisfy the third condition. Please analyze why the current set of assertions cannot prove the postcondition. The C program to be verified is shown as follows:
		\{Program\}
		\\
		\\
		$\#\#\#$ Require $\#\#\#$
		
		Directly provide a concise conclusion that helps identify the missing assertions. Do not repeat the problem description.\label{prompt: reason failure}
	\end{tcolorbox}
\end{mybox}

\subsubsection{2-CNF Lemma Synthesizer}\label{sec: 2-CNF Lemma Learner}

This section introduces the 2-CNF Lemma Synthesizer, which searches for candidate lemmas from the combinatorial space of literals. The overall algorithm is summarized in Algorithm~\ref{alg: 2CNFLemmaLearner}. To perform an ordered and manageable search, we use the grammatical form of conjunctive normal form(CNF). At the same time, this grammatical form is naturally suitable for lemma search and will not cause redundancy within the lemma sequence. The CNF clauses are all stored in the \texttt{clauseSet}, including unit clauses consisting of a single literal and binary clauses consisting of the disjunction of two literals. To ensure search efficiency, we do not combine clauses consisting of more than two literals. In addition, the clauses in \texttt{clauseSet} are labeled with whether they are legal or not. We define illegal clauses as follows:

\begin{definition}[Illegal clause]\label{defi: illegal clause}
	An illegal clause refers to any clause that has been previously prohibited or appears as a component of any lemma in \texttt{lemmaSeq}, which would cause redundancy or failure in the candidate lemma.
\end{definition}

In each search, we maintain a temporary data structure \texttt{lemmaList}, which stores all CNFs to be checked. At the beginning of the search, the CNFs of \texttt{reserveSet} are transferred to \texttt{lemmaList} as the initial elements. If new literals are received, we update the \texttt{clauseSet}. First, the literals become new unit clauses. Then, we produce new binary clauses by conducting pairwise disjunction between different literals within new literals and between each new literal and each old unit clause. These new clauses will be added to \texttt{clauseSet} and \texttt{lemmaList}.

Next, we search for candidate lemmas in \texttt{lemmaList}. For each CNF, if it is falsified by any positive counterexample, we remove it from \texttt{lemmaList} and prohibit it from being conjuncted with other clauses. If it is falsified by any inductive counterexample, we decide whether to refine it based on its clause number. If it contains only one clause, we refine it by conducting a conjunction between it and each legal unit clause. Otherwise, we just remove it from \texttt{lemmaList}. If it cannot exclude any negative counterexample, we move it from \texttt{lemmaList} to \texttt{reserveSet} because it may exclude a negative counterexample in the future. After the search procedure, if \texttt{lemmaList} becomes an empty set, we stop searching and go to the fallback learning process. Otherwise, the shortest one will be selected as a candidate lemma and submitted to the Lemma Validator for a rigorous check.

%In our design, the 2-CNF search strategy is a trade-off between expressive power and search efficiency. The 2-CNF structure constrains the combinatorial form of lemmas, allowing efficient and controllable search. Expressive power, on the other hand, is primarily provided by literals generated by LLMs, which can encode rich semantic information. Empirical evidence shows that composing literals using 2-CNF is sufficient to synthesize the majority of loop invariants in the testing benchmarks. Moreover, in an incremental setting, 2-CNF–based search strategies exhibit more stable and controllable behavior than monolithic approaches.

In our design, the 2-CNF search strategy is a trade-off between expressive power and search efficiency. The 2-CNF structure constrains the combinatorial form of lemmas to enable efficient search, while expressive power is provided primarily by literals generated by LLM. Our empirical results indicate that 2-CNF combinations of literals generated by the LLM are sufficient to synthesize the majority of loop invariants in the evaluated benchmarks. Moreover, in an incremental setting, the performance of 2-CNF–based search strategies is more stable and controllable than monolithic approaches.

% Our literals are mostly generated by LLMs and have sufficient expressive power. With the introduction of new literals, the expressive power of 2-CNF gradually increases. Empirical evidence shows that composing literals using 2-CNF is sufficient to synthesize the majority of loop invariants, and that in an incremental setting, search strategies based on 2-CNF are more stable and controllable than the monolithic approach.

%Here, the 2-CNF search strategy is a trade-off between expressive power and search efficiency. Our literals are mostly generated by LLMs and have sufficient expressive power. With the introduction of new literals, the expressive power of 2-CNF gradually increases. Empirical evidence shows that composing literals using 2-CNF is sufficient to synthesize the majority of loop invariants, and that in an incremental setting, search strategies based on 2-CNF are more stable and controllable than the monolithic approach.

\begin{algorithm}[t]
\caption{Procedure \textsc{2CNFLemmaSynthesizer}}
\label{alg: 2CNFLemmaLearner}
\KwIn{New literals $\mathit{newLiterals}$, clause set $\mathit{clauseSet}$, lemma sequence $\mathit{lemmaSeq}$, counterexample set $\mathit{CESet}:(\mathit{CE_p},\mathit{CE_i},\mathit{CE_n})$, reserved lemmas $\mathit{reserveSet}$.}
\KwOut{Candidate lemma $\mathit{candidate}$}

\Fn{\FCNFLemmaSynthesizer{$\mathit{newLiterals}, \mathit{clauseSet}, \mathit{lemmaSeq}, \mathit{CESet},\mathit{reserveSet}$}}{
    $\mathit{lemmaList}, candidate\gets \emptyset$\;
    \If{$\mathit{newLiterals}\not = \emptyset$}{
        $\mathit{oldLiterals}\leftarrow GetUnitClause(\mathit{clauses})$\;
        $\mathit{newClauses}\leftarrow LogicalDisjunction(\mathit{newLiterals},\mathit{oldLiterals})$\;
        $\mathit{clauses}.append(\mathit{newClauses})$\;
        $\mathit{lemmaList}\leftarrow\mathit{newClauses}$\;
    }
    $SetMove(\mathit{reserveSet},\mathit{lemmaList})$\tcp*{Move the elements of $\mathit{reserveSet}$ to $\mathit{lemmaList}$}
    \ForEach{$\mathit{CNF} \in \mathit{lemmaList}$}{
        \If{$not \ CheckAll(\mathit{CNF},\mathit{CE_p})$}{
            $Prohibit(\mathit{clauseSet},\mathit{CNF})$\;
        }
        \ElseIf{$not \ CheckAll(\mathit{CNF},\mathit{CE_i})$}{
            $\mathit{lemmaList}.UnitRefine(\mathit{CNF},\mathit{clauseSet})$\;
        }
        \ElseIf{$not \ CheckAny(\mathit{CNF},\mathit{CE_n})$}{
            $SingleMove(\mathit{CNF},\mathit{lemmaList},\mathit{reserveSet})$\;
        }
    }
    \If{$\mathit{lemmaList}\not = \emptyset$}{
        $\mathit{candidate}\gets GetShortest(\mathit{lemmaList})$\;
        $\mathit{reserveSet}.append(\mathit{lemmaList} \ / \ \mathit{candidate})$\;
    }    
    \Return $\mathit{candidate}$ \;
}
\end{algorithm}

\subsubsection{Lemma Validator}

This section introduces the Lemma Validator, which is designed to validate the candidate lemma with respect to the \texttt{lemmaSeq} and guide the subsequent search process. In practice, this module is responsible for validating the candidates proposed by both synthesizers. However, the validator described in this section is only used to validate the results of the 2-CNF Lemma Synthesizer. The validation of ICE-DT will be introduced in Section~\ref{sec: ICE-DT Fallback Learner}. When the candidate does not hold initially or is not inductive relative to the \texttt{lemmaSeq}, the processing methods are similar to the search process in Section~\ref{sec: 2-CNF Lemma Learner}, except for the degree of refinement. Here, we refine the candidate using all legal clauses and do not restrict the number of clauses because the high-quality candidate deserves more attention. When the candidate lemma is proven to be initial and relatively inductive, we add it to the \texttt{lemmaSeq}. If the accumulated lemmas are enough to prove the property, the conjunction of the \texttt{lemmaSeq} will be returned as a complete loop invariant. Otherwise, we use it to filter the current counterexamples as described in Section~\ref{section: Incremental ICE}. At this point, due to the introduction of a new valid lemma, all legal clauses may become relatively inductive again. Thus, we refresh the search procedure by replacing \texttt{reserveSet} with all legal clauses in \texttt{clauseSet}. 

\begin{algorithm}[t]
\caption{Procedure \textsc{LemmaValidator}}
% \vspace{-1em}
\label{alg:LemmaValidator}
\KwIn{Candidate Lemma $\mathit{candidate}$, clause set $\mathit{clauseSet}$, lemma sequence $\mathit{lemmaSeq}$, counterexample set $\mathit{CESet}$, reserved lemmas $\mathit{reserveSet}$.}
\KwOut{Proof status $\mathit{status}$}

\Fn{\FLemmaValidator{$\mathit{candidate}, \mathit{clauseSet}, \mathit{lemmaSeq}, \mathit{CESet},\mathit{reserveSet}$}}{
    $[\mathit{Pre},\mathit{Indu},\mathit{Post}], \mathit{newCE} \leftarrow SMTCheck(\mathit{candidate},\mathit{lemmaSeq})$\;
    $\mathit{CESet}.append(\mathit{newCE})$\;
    \If{$not \ \mathit{Pre}$}{
        $Prohibit(\mathit{clauseSet},\mathit{candidate})$\;
    }
    \ElseIf{$not \ \mathit{Indu}$}{
        $refineSet\gets AllRefine(\mathit{candidate},\mathit{clauseSet})$\;
        $\mathit{reserveSet}.append(refineSet)$\;
    }
    \ElseIf{$not \ \mathit{Post}$}{
        $\mathit{lemmaSeq}.append(\mathit{candidate})$\;    
        $CEFilter(\mathit{CESet},\mathit{candidate})$\;
        $\mathit{reserveSet}\leftarrow GetLegal(\mathit{clauses})\qquad$\tcp{Refresh}
    }
    \Else{
        $\mathit{lemmaSeq}.append(\mathit{candidate})$\; 
        \Return \textbf{True} \;                
    }
    \Return \textbf{False} \;
}
\end{algorithm}

\subsection{ICE-DT Fallback Learning Process}\label{sec: ICE-DT Fallback Learner}

When the 2-CNF Lemma Synthesizer fails to propose a new candidate lemma, we turn to ICE-DT~\cite{icedt} to synthesize the missing part. Since our Incremental ICE framework involves a counterexample filtering mechanism, the retained counterexamples partially characterize the properties that the missing lemma must satisfy. ICE-DT can perform fine-grained synthesis with respect to the full set of counterexamples, whereas LLMs are not well-suited for this task. Thus, here we iteratively call ICE-DT until it synthesizes the final lemmas or reaches the maximum number of iterations. In practice, we observe that ICE-DT either produces incorrect predicates or complete invariants, and only rarely produces lemmas that are insufficient to prove the postcondition. This is because satisfying inductiveness is the core challenge of ICE-DT. Therefore, when validating the candidates of ICE-DT, we only check whether they can fully solve the problem, and do not refine the candidates or add them to \texttt{reserveSet}. Integrating with ICE-DT can alleviate the LLM's lack of precise reasoning ability and improve the loop invariant synthesis efficiency.

\section{Evaluation}\label{sec: evaluation}

In this section, we present the experimental results that are designed to answer the following research questions:

\vspace{0.1cm}

\noindent\textbf{RQ1.} How does LimICE compare in efficiency and effectiveness to the state-of-the-art methods?\\
\textbf{RQ2.} How does the choice of different LLM backends affect the performance of LLM-based methods? \\
\textbf{RQ3.} How does the incremental combination compare to the monolithic combination?\\
\textbf{RQ4.} How do different design choices within LimICE affect its overall performance?

\subsection{Setup}

\indent\textit{Benchmarks.} We evaluate our approach on both linear and non-linear benchmarks to cover complementary problem characteristics. The linear benchmarks comprise 367 instances drawn from three subsets: 313 problems collected by LaM4Inv~\cite{lam4inv}, including 133 instances originally introduced by Code2Inv~\cite{code2inv} and 180 new benchmarks newly constructed in LaM4Inv, as well as 54 new benchmarks crafted by us. The new benchmarks are obtained by manually applying equivalence-preserving rewrites of the benchmarks in the CHC-COMP repository~\cite{chccomp}. For existing linear benchmark sets, we discover some systemic errors and apply a set of correctness-oriented and semantics-preserving refinements to ensure benchmark validity, including removing redundant classes, reforming trivial instances, and eliminating factual inconsistencies; as a result, the refined benchmarks become slightly more challenging due to the removal of trivial or redundant cases. The non-linear benchmarks consist of 50 instances collected by Clause2Inv~\cite{clause2inv}, including 30 instances originally introduced by LIPuS~\cite{lipus} and 20 benchmarks newly constructed in Clause2Inv. These benchmarks have also undergone some error corrections. Detailed descriptions of the benchmarks and all modifications are provided in the supplementary material.

% To comprehensively evaluate the efficiency and effectiveness of the different methods, we prepared a dataset including 367 linear problems and 50 nonlinear problems. The linear dataset consists of three parts: 133 problems collected by Code2Inv~\cite{code2inv}, 180 problems collected by LaM4Inv~\cite{lam4inv} and 54 new benchmarks crafted by us. The second part originally contained 183 instances. During our reproduction process, we identified several systematic issues and corrected them. As part of this process, we excluded three instances that could not be reliably fixed, resulting in a final benchmark of 180 instances. In addition, we manually crafted 54 new benchmarks based on the 2025 CHC-COMP, which are more challenging than those in the first two parts. The nonlinear dataset consists of two parts: 30 problems collected by LIPuS~\cite{lipus} and 20 problems collected by Clause2Inv~\cite{clause2inv}. More details of the benchmark construction are included in Appendix \ref{sec: Benchmark Details}.

\noindent\textit{Baselines.} To evaluate our approach, we compare it with the following baselines.

\begin{itemize}
    \item ICE-DT~\cite{icedt} synthesizes program invariants by extending decision tree learning to the ICE framework, enabling consistent and convergent learning from counterexamples;
    \item ICE-DT-Interval~\cite{icedtinterval} accelerates loop invariant learning by introducing interval counterexamples and extending decision tree learning to handle them, significantly reducing learning rounds and verification time;
    \item LoopInvGen~\cite{loopinvgen} synthesizes loop invariants by iteratively using data-driven precondition synthesis with on-demand feature learning and SMT-based counterexample refinement;
    \item Code2Inv~\cite{code2inv} synthesizes loop invariants using a reinforcement-learning approach guided by graph neural network representations of programs and feedback from a theorem prover;
    \item LIPuS~\cite{lipus} synthesizes loop invariants by generating invariant templates using reinforcement learning and then determining the template parameters by solving counterexamples;
    \item CLN2INV~\cite{cln2inv} synthesizes loop invariants by learning explicit SMT formulas from execution traces using Continuous Logic Networks, which map logical constraints to differentiable continuous semantics;
    \item LaM4Inv~\cite{lam4inv} synthesizes loop invariants by filtering valid components of candidate loop invariants generated by LLM via bounded model checking and reassembling them into a new invariant;
    \item Clause2Inv~\cite{clause2inv} synthesizes loop invariants by generating atomic literals using LLMs and combining them into a correct invariant using a counterexample-driven method under a generate-combine-check framework.
\end{itemize}

Among these baselines, LaM4Inv and Clause2Inv are LLM-based methods, LoopInvGen is a traditional symbolic method, and the rest are learning-based methods. For linear benchmarks, we compare all baselines except LIPuS, while for nonlinear benchmarks, we compare LIPuS, LaM4Inv, Clause2Inv, and ICE-DT-Interval, which are good at handling nonlinear problems.

\textit{Evaluation metrics.} To evaluate the efficiency and effectiveness of different methods, we record the number of successfully generated loop invariants and the average time consumed on all benchmarks successfully solved. In some experiments, we also compared the number of SMT solver queries to aid in efficiency comparison.

\textit{Implementations.} All experiments are conducted under identical settings. In terms of hardware, the CPU is the 13th Gen Intel(R) Xeon(R) Silver 4314, equipped with 256GB of onboard RAM. The GPU is RTX A6000. For the LLM-based method, we all adopt DeepSeek V3.2 by default. For LimICE, we set the iteration number of ICE-DT in each iteration to 30. We set a time limit of 600 seconds for each method to solve per problem.

\subsection{RQ1. Efficiency and Effectiveness}\label{sec: rq1}

\begin{table*}[t]
\centering
\caption{Performance Comparison on Linear Problems.}
\label{tab:linear comparison}
\begin{tabular}{lrrrrrrrr}
\toprule
\multirow{2}{*}[-0.3ex]{Methods} & \multicolumn{2}{c}{NeurIPS-18 (133)}                    & \multicolumn{2}{c}{ASE-24 (180)}                        & \multicolumn{2}{c}{CHC-Comp (54)}                      & \multicolumn{2}{c}{Total (367)}                          \\
\cmidrule(lr){2-3}
\cmidrule(lr){4-5}
\cmidrule(lr){6-7}
\cmidrule(lr){8-9}
                  & Solved             & Time (s)                & Solved             & Time (s)                & Solved             & Time (s)                 & Solved             & Time (s)                 \\
\midrule
\textbf{LimICE}            & \textbf{133} & \textbf{3.9} & \textbf{179} & \textbf{9.1} & \textbf{37} & \textbf{85.1} & \textbf{349} & \textbf{15.2} \\
LoopInvGen        & 103 &  6.4 & 104 & 12.8 &  0 &   0.0 & 207 &  9.6 \\
ICE-DT            & 128 &  2.8 & 111 &  6.0 &  8 &  76.1 & 247 &  6.6 \\
ICE-DT-Interval   & 128 &  0.8 & 123 &  4.3 & 12 &  25.2 & 263 &  3.5 \\
Code2Inv          & 106 & 46.5 &  85 & 71.6 &  4 & 149.2 & 195 & 59.5 \\
CLN2INV           & 120 &  0.2 &  94 &  0.2 &  2 &   0.3 & 216 &  0.2 \\
Clause2Inv        & 119 & 13.0 & 170 & 20.0 & 22 & 109.6 & 311 & 23.7 \\
LaM4Inv           & 111 & 18.9 & 163 & 29.5 & 18 &  70.0 & 292 & 28.0 \\        
\bottomrule
\end{tabular}
\end{table*}

\noindent \textit{Performance comparison on linear problems} We first compare the different methods' performance on linear benchmarks. The results are shown in Table~\ref{tab:linear comparison}, where the time column is the average time computed on all benchmarks successfully solved by the tool. It can be observed that LimICE achieves the best results on all datasets in terms of the number of solved benchmarks.

LimICE generates at least 86 more invariants than the non-LLM baseline. In comparison, existing non-LLM methods solve substantially fewer instances. Although these methods achieve relatively strong performance on the first, less complex dataset, their effectiveness degrades noticeably as the dataset complexity increases. In particular, on the newly introduced dataset, even the best-performing baseline solves only 22$\%$ of the instances. LimICE integrates LLM with ICE-DT under an incremental framework and exhibits good scalability. However, the average solving time of LimICE is higher than that of most of these methods. This is primarily due to the additional communication overhead introduced by querying the LLM, as well as the fact that our approach solves more challenging problems.

Compared with the LLM-based method, LimICE demonstrates higher efficiency and effectiveness. It takes 64$\%$ of the time of Clause2Inv while solving 38 more problems; likewise, it takes 54$\%$ of the time of LaM4Inv while solving 57 more problems. There is no model advantage here because all three methods adopt DeepSeek V3.2. But LimICE incorporates a warm start mechanism and the ICE-DT fallback mechanism, and these components allow it to move beyond solely relying on LLM to solve problems. In contrast, both LaM4Inv and Clause2Inv are limited by the efficiency and capability of LLMs.

\begin{table*}[h]
\centering
\caption{Performance Comparison on Nonlinear Problems.}
\label{tab:nonlinear comparison}
\begin{tabular}{lrrrrrr}
\toprule
\multirow{2}{*}[-0.3ex]{Methods} & \multicolumn{2}{c}{ISSTA-23 (30)}     & \multicolumn{2}{c}{ISSTA-25 (20)}     & \multicolumn{2}{c}{Total (50)}        \\
\cmidrule(lr){2-3}
\cmidrule(lr){4-5}
\cmidrule(lr){6-7}

                  & \multicolumn{1}{r}{Solved} & Time (s) & \multicolumn{1}{r}{Solved} & Time (s) & \multicolumn{1}{r}{Solved} & Time (s) \\ 
\midrule
\textbf{LimICE}            & \textbf{29} & \textbf{10.0}  & \textbf{18} & \textbf{6.8} & \textbf{47} & \textbf{8.8} \\ 
LIPuS             & 20 &   6.3 &  0 &     0 & 20 &   6.3  \\
ICE-DT-Interval   &  6 & 188.1 & 10 & 136.5 & 16 & 155.9  \\ 
Clause2Inv        & 23 &  33.7 & 15 &   7.9 & 38 &  23.5   \\ 
LaM4Inv           & 16 &  38.6 & 15 &  43.0 & 31 &  40.7   \\ 
\bottomrule
\end{tabular}
\end{table*}

\noindent\textit{Performance comparison on nonlinear problems} In this paper, we compare LimICE with LIPuS, ICE-DT-Interval, Clause2Inv, and LaM4Inv, because these methods perform well on nonlinear problems. The results are shown in Table~\ref{tab:nonlinear comparison}. It can be observed that LimICE has significant advantages in both effectiveness and efficiency. 

For ICE-DT-Interval, its time efficiency differs significantly between linear and nonlinear benchmarks, with the average time increasing from 3.5s to 155.9s. One possible explanation for this phenomenon is that it can only use linear expressions when constructing the decision tree, and this does not solve nonlinear problems well. However, LLM-based methods are not limited by expressive ability and thus exhibit more stable performance. For LIPuS, its two-dimensional reward design relies on manually determined parameters and leads to significant performance fluctuations across different datasets.

Compared with the other two LLM-based methods, LimICE has a significant efficiency advantage. Through careful analysis of the synthesis process, we find that the incremental combination strategy plays an important role. The goal of LimICE is to find new lemmas instead of a complete loop invariant, and this has an advantage in exploring the space of literal combinations. So, LimICE can solve 37 benchmarks in a single LLM query round. In contrast, there are several times that all the literals that make up a correct invariant have been generated, but Clause2inv fails to combine a complete loop invariant. In addition, LaM4Inv’s strategy of filtering valid predicates from invariants appears to be less effective for solving non-linear problems.

% , on the other hand, generally requires more queries to the LLM, resulting in greater communication overhead.

\subsection{RQ2. Different LLM}

In this section, we evaluate the adaptability of LLM-based approaches across different LLM backends. Since DeepSeek has been used for comparison in Section~\ref{sec: rq1}, we further consider two representative models: GPT-3.5-Turbo and QWen2.5-7B-Instruct. GPT-3.5-Turbo serves as a widely adopted baseline model, while QWen2.5-7B-Instruct is chosen to assess the effectiveness of our approach on models with significantly fewer parameters. In order to strike a balance between the communication overhead and parallel response costs associated with local and online LLMs, this experiment also constrains the maximum number of iterations to 50 in addition to the time limit. The results are shown in Table~\ref{tab: different LLMs}. It can be observed that our approach provides stability in linear problems. Even with a relatively small 7B model, LimICE still solves 88.8$\%$ of the linear benchmarks. This is because the integration of the ICE-DT fallback mechanism allows our method to avoid being overly dependent on the capabilities of LLMs. The role of fallback mechanisms becomes increasingly prominent as model capabilities decline.

However, we observe that non-linear problems are more challenging when using the 7B model. This is because the ICE-DT can only generate decision trees using linear decision nodes, which are inherently less effective for handling non-linear problems. Thus, only after the LLM solves most of the problems can ICE-DT synthesize the final lemmas. Despite this limitation, our approach still solves more problems with higher efficiency than both LaM4Inv and Clause2Inv. We also observe that Clause2Inv exhibits reasonable stability on larger models. However, it is limited by the capabilities of the models, leading to significant performance degradation on the 7B model. In addition, LaM4Inv is most sensitive to the choice of model, as it relies on filtering valid predicates directly from the complete loop invariants generated by LLMs.

\begin{table*}[t]
	\centering
    \begin{threeparttable}
	\caption{Performance Comparison of LLM-Based Methods on Different LLM.}
	\label{tab: different LLMs}
	\setlength{\tabcolsep}{4pt}
	\begin{tabular}{cccccccc}
		\toprule
		\multirow{2}{*}[-0.3ex]{Models}                                                       & \multirow{2}{*}[-0.3ex]{Linearity} & \multicolumn{3}{c}{Solved Benchmarks}    & \multicolumn{3}{c}{Average Time (s)}  \\
		\cmidrule(r){3-5}
		\cmidrule(r){6-8}
		&                                    & LimICE       & LaM4Inv & Clause2Inv & LimICE        & LaM4Inv & Clause2Inv       \\
		\midrule
		\multirow{2}{*}{QWen2-5:7B} & Linear & \textbf{326 (39)} & 221     & 258        & \textbf{34.4} & 159.3   & 91.4        \\
		& Nonlinear                          & \textbf{32 (0)}   & 10      & 28         & \textbf{45.6} & 126.4   & 115.9       \\
		\midrule
		\multirow{2}{*}{GPT 3.5 Turbo} & Linear & \textbf{339 (13)} & 306     & 306     & \textbf{15.1} & 60.6    & 25.9        \\
		& Nonlinear                          & \textbf{45 (1)}   & 20      & 37         & \textbf{26.4} & 92.3    & 29.8        \\   
		\bottomrule	
	\end{tabular}

    \begin{tablenotes}
    \footnotesize
    \item Notice: In the "LimICE" column of the Solved Benchmarks, the number in parentheses indicates the number of problems directly addressed by the ICE-DT fallback synthesizer.
    \end{tablenotes}
    \end{threeparttable}
\end{table*}

\subsection{RQ3. Incremental Combination vs. Monolithic Combination}\label{sec: RQ3 Incremental vs Monolithic}

In this subsection, we remove all components and simply compare the ability of LimICE and Clause2Inv to combine invariants from the atomic literal set, with the latter being an excellent instance of the monolithic method. To control the randomness of LLM-based literal generation and ensure fair comparison across methods, we let DeepSeek generate 10 consecutive rounds of literals using the random generation method and repeat this process five times to obtain five fixed generation sequences. Then, we use these fixed sequences as the input of two methods and record, for each round, the number of problems successfully solved and the number of SMT solver calls incurred. The final results are summarized in Table~\ref{tab: incremental vs monolithic}. Success@B denotes the average performance in round B over the five sequences. It can be observed that the incremental combination strategy is superior to the monolithic combination strategy in both effectiveness and efficiency.

The overall scope reports the results on the full benchmark suite. For linear problems, our incremental approach performs slightly better than the monolithic approach. However, for nonlinear problems, we demonstrate a relatively significant advantage. Under identical inputs and over 10 iterations, our incremental method solves 27.8$\%$ more problems without requiring additional SMT solver calls, and consistently outperforms the monolithic approach in every round. This shows that the incremental approach offers better performance and stability.

In addition, through careful analysis of the results, we found that the majority of benchmarks can be quickly solved by two methods within a few rounds. Such simple benchmarks tend to obscure the differences among methods. Therefore, we filter out benchmarks that can be solved by both methods within five rounds across all five fixed sequences, resulting in a set of challenging benchmarks including 105 linear problems and 24 nonlinear problems. The results on these benchmarks are shown in the hard scope, and it can be observed that there is a more obvious difference between the two approaches, and the incremental approach performs better on complex benchmarks.

\begin{table*}[t]
  \caption{Performance Comparison Between the Incremental Approach and the Monolithic Approach.}
  \label{tab: incremental vs monolithic}
  \centering
  \setlength{\tabcolsep}{4pt}
  \begin{tabular}{p{1.2cm}llrrrrrrrr}
    \toprule
    \multirow{2}{*}[-0.3ex]{Scope} &
    \multirow{2}{*}[-0.3ex]{Linearity}  &
    \multirow{2}{*}[-0.3ex]{Methods}     &
    \multicolumn{2}{c}{Success@1} &
    \multicolumn{2}{c}{Success@2} &
    \multicolumn{2}{c}{Success@5} &
    \multicolumn{2}{c}{Success@10} \\    
    \cmidrule(r){4-5}
    \cmidrule(r){6-7}
    \cmidrule(r){8-9}
    \cmidrule(r){10-11}
    & & &
    Solved & SMT &
    Solved & SMT &
    Solved & SMT &
    Solved & SMT \\
    \midrule
    \multirow{4}{*}[-0.3ex]{Overall} & \multirow{2}{*}{Linear}
      & Monolithic    & 254 & 4.0 & 271  & 4.3 & 289 & 4.7 & 298 & 4.9 \\
    & & Incremental   & 257 & 3.8 & 279  & 4.1 & 298 & 4.5 & 309 & 4.7 \\
    \cmidrule(r){2-11}
     & \multirow{2}{*}{Nonlinear}
      & Monolithic    & 27  & 6.7 & 31  & 7.6 & 33  & 8.9 & 36 & 10.1 \\
    & & Incremental   & 36  & 7.1 & 39  & 7.6 & 42  & 8.7 & 46 & 9.9 \\
    \midrule
    \multirow{4}{*}[-0.3ex]{Hard} & \multirow{2}{*}{Linear}
      & Monolithic    & 9   & 7.5 & 18  & 8.6 & 27  & 9.5 & 36 & 10.0 \\
    & & Incremental   & 12  & 7.3 & 24  & 8.0 & 36  & 9.3 & 47 & 9.7 \\
    \cmidrule(r){2-11}
     & \multirow{2}{*}{Nonlinear}
      & Monolithic    & 3   & 10.1 & 5  & 12.0 & 7  & 14.7 & 10 & 17.6 \\
    & & Incremental   & 10  & 10.7 & 13 & 11.3 & 16 & 13.2 & 20 & 15.2 \\
    \bottomrule
  \end{tabular}
\end{table*}

\subsection{RQ4. Ablation Study}

We have known that the incremental combination strategy can solve the majority of nonlinear problems, as shown in section~\ref{sec: RQ3 Incremental vs Monolithic}. So, only linear problems will be used to compare the performance of different models in the ablation study. Meanwhile, the hard benchmarks identified in section~\ref{sec: RQ3 Incremental vs Monolithic} are also used here for a more comprehensive analysis. The results are shown in Table~\ref{tab: ablation}.

\noindent\textit{LLM Only.} The open-source tool of LaM4Inv~\cite{lam4inv} provides a dedicated configuration that allows us to ask LLM to generate complete loop invariants directly. We use it to test the effect of using DeepSeek alone. The result is shown in Table~\ref{tab: ablation}, and we can find that performance declines significantly, with the number of solved problems decreasing from 349 to 195, while the average time increased from 15.2 seconds to 37.1 seconds. This means that when dealing with complex problems, it is necessary to take additional measures to strengthen the ability of LLMs.

\noindent\textit{Effect of different components.} We next analyze the impact of individual components, including the reason generation method, the warm start mechanism, and the ICE-DT fallback mechanism. The LimICE-base consists only of the random generation method, the 2-CNF Lemma Synthesizer, and the Lemma Validator. Through comparison with Section~\ref{sec: rq1}, we observe that even the LimICE-base configuration can achieve SOTA performance. We then incrementally add each component to the LimICE-base and report the corresponding results. The results show that all three methods can help solve difficult problems. Using the warm start and ICE-DT fallback mechanism can solve more problems while significantly improving solution efficiency. The combination of reason generation methods can allow LLM to generate more targeted results and mitigate the problem of model output saturation. This approach fully exploits the LLM’s capabilities, but its efficiency gains remain limited. The LimICE-full, which integrates all three components, achieves the best performance, suggesting that these components complement each other.

\begin{table*}[t]
  \caption{Result of Ablation Study.}
  \label{tab: ablation}
  \centering
\begin{tabular}{ccccc}
\toprule
\multirow{2}{*}[-0.3ex]{Models} & \multicolumn{2}{c}{Overall(133/180/54)} & \multicolumn{2}{c}{Hard(24/35/46)} \\
\cmidrule(lr){2-3} 
\cmidrule(lr){4-5}
                                            & Solved                 & Time (s)       & Solved             & Time (s)      \\
\midrule
LLM Only                                    & 195(70/117/8)          & 37.1           & 21(2/13/6)         & 67.0          \\
LimICE-base                                & 322(122/175/25)        & 21.0           & 63(16/30/17)       & 84.9          \\
LimICE-base + Reason                       & 333(125/177/31)        & 21.5           & 71(16/32/23)       & 75.8          \\
LimICE-base + Warm                         & 342(133/178/31)        & 14.6           & 80(24/33/23)       & 45.1          \\
LimICE-base + ICEDT                        & 339(133/177/29)        & 12.1           & 77(24/32/21)       & 34.6          \\
LimICE-full                                & 349(133/179/37)        & 15.2           & 87(24/34/29)       & 46.1          \\
\bottomrule
\end{tabular}
\end{table*}

\subsection{Case Studies}

\subsubsection{Case study 1} We first show a complex problem that is only solved by LimICE. The problem is shown in Figure~\ref{fig: complex linear case}. 

\begin{figure}[h]
	\centering
	\begin{minipage}{0.42\textwidth}
		\begin{lstlisting}[language=C]
			int main(){
				int x, y;
				// pre-conditions
				x = 0;
				y = 0;
				// loop body
				while(unknown()){
					if(x >= 7500){
						if(x >= 12500){
							y = y - 2;
						}else{
							y = y + 1;
						}
					}
				\end{lstlisting}
			\end{minipage}
			\hfill
			\begin{minipage}{0.42\textwidth}
				\begin{lstlisting}[language=C]
					else{
						if(x >= 2500){
							y = y + 1;
						}else{
							y = y - 2;
						}
					}
					x = x + 1;
				}
				// post-condition
				if(x == 15000){
					assert(y == 0);
				}
			}
		\end{lstlisting}
	\end{minipage}
	\caption{A complex benchmark of cast study 1.}
	\label{fig: complex linear case}
\end{figure}

The final loop invariant synthesized by LimICE is as follows:

\begin{equation*}
\begin{matrix}
x >= 0 \ \land \ ( (y >= -2 * (x - 2500) \ \lor \ y >= -10000) \ \land \ y >= -2 * x ) \ \land \\
 (y <= (x - 2500) \ \lor \ y = -2 * (x - 2500) - 5000) \ \land \\
 (y <= (x - 7500) \ \lor \ y = -2 * (x - 2500) - 5000) \ \land \\ 
y <= -2 * (x - 12500) + 1 * 5000 + 1 * 2500 \ \land \\
 2*x + y <= 30000 \ \land \ (7500 - x >= 0 - y \ \lor \ y = -2 * x + 30000).
\end{matrix}
\end{equation*}

The complete invariant is complex, and when the problem is solved, the number of cumulative literals generated by LLM and warm start has reached 91. Finding such an invariant in the vast combinatorial space is difficult. However, LimICE adopts an incremental approach; it discovers seven relatively simple lemmas during the synthesis process rather than trying to obtain the complete invariant once. Therefore, it has certain advantages when facing complex problems.

\subsubsection{Case study 2} Next, we show another complex example to demonstrate how the components of our system interact and collectively contribute to solving challenging cases. The problem is shown in Figure~\ref{fig: advantage benchmark}.

\begin{figure}[h]
	\centering
	\begin{minipage}{0.42\textwidth}
		\begin{lstlisting}[language=C]
			int main(){
				int x, y, z;
				// pre-conditions
				x = 0;
				y = 10;
				z = 0;
				// loop body
				while(unknown()){
					if(x == y){
						z = 0;
					}
				\end{lstlisting}
			\end{minipage}
			\hfill
			\begin{minipage}{0.42\textwidth}
				\begin{lstlisting}[language=C]
					else{
						z = z + 1;
					} 
					x = (x + 1) % 10;
					y = (y - 1) % 10;
				}
				// post-condition
				assert(z <= 5);
			}
		\end{lstlisting}
	\end{minipage}
	\caption{A complex benchmark of cast study 2.}
	\label{fig: advantage benchmark}
\end{figure}

LimICE first synthesizes the following five lemmas:

\begin{equation*}
\begin{array}{ll}
(1)\; x \ge 0 \;\lor\; x = 0
& (4)\;  (x + y = 10 \;\lor\; x = y)
              \;\land\;
              (x + y = 10 \;\lor\; x = 0)
\\[3pt]
(2)\; x \le 9
& (5)\; z \le 10 - (y - x) \;\lor\; x = y
\\[3pt]
(3)\; y \ge 0
&
\end{array}.
\end{equation*}

In the literals that make up the lemmas, $x=y$ and $x=0$ are derived from the warm start. Based on the five lemmas above, the ICE-DT fallback mechanism successfully synthesizes the final lemmas:

\begin{equation*}
    ( y + z <= 10 \ \land \ y <= 4 \ \land \ y + z <= 4 ) \ \lor \ ( y + z <= 10 \ \land \ y > 4 ).
\end{equation*}

These lemmas can solve the problem, and the whole process consumes only 15.1 seconds. The absence of any component will make the problem unsolvable or significantly increase the time cost.

\section{Threats to Validity} \label{sec: discussion}

There are three main validity threats to the validity of our work. First, the LLM is an important component of our method, which introduces inherent nondeterminism. However, techniques warm start, and the ICE-DT fallback mechanism in our design reduces reliance on the LLM and helps improve overall stability. In addition, we also conduct experiments with fixed LLM outputs to more accurately evaluate the effectiveness of our method.

Second, although we evaluated our approach on a set of 417 benchmarks, the complexity of these benchmarks is generally lower than that of real-world programs. However, the selected benchmarks are widely used in the research community and originate from internationally recognized competitions, making them representative of common evaluation scenarios. Future work will develop more complex and comprehensive benchmarks.

Third, the capabilities of SMT solvers limit our method. It is an essential component for ensuring the correctness of the results, but its efficiency is limited when performing operations such as multiplication, division, and exponentiation. In addition, recent research~\cite{smtvalid} has found some soundness bugs in the SMT solver, which might threaten the verification results.

\section{Related Work} \label{sec: related work}

Traditional methods for invariant generation include techniques such as abstract interpretation~\cite{abstract1,abstract2,abstract3}, counterexample guided abstraction refinement~\cite{cerefine1,cerefine2}, model checking~\cite{modelcheck1,modelcheck2}, Craig interpolation~\cite{craig1,craig2}, constraint solving~\cite{constraint1,constraint2}, abductive reasoning~\cite{abductive,abductive2}, and dynamic inference~\cite{dynamic1,dynamic2,dynamic3}. In traditional methods, IC3~\cite{ic31,ic32,ic33} applies the relative inductiveness to automated synthesis, which serves as the inspiration for our approach. These methods are generally stable on different problems, but face scalability issues.

Learning-based methods typically follow a guess-and-check framework, leveraging machine learning techniques to generate candidate invariants from counterexample sets, program execution sequences, or program code. Garg et al.~\cite{ice} introduce the inductive counterexample and propose ICE, which is a robust and strongly convergent framework for counterexample-based learning methods. Variants of the ICE framework include ICE-DT~\cite{icedt}, ICE-DT-Interval~\cite{icedtinterval}, Hore-ICE~\cite{hornice}, Tapis~\cite{icearray}, etc. CLN2INV~\cite{cln2inv} and G-CLN~\cite{gcln} learn an invariant from a program execution sequence using a continuous logic network. Code2Inv~\cite{code2inv} and LIPuS~\cite{lipus}, on the other hand, utilize a reinforcement learning framework to learn invariants directly from the programs.

Recent research has tried to leverage LLMs to synthesize loop invariants. Akhond et al.~\cite{llmEvaluate} conduct a detailed and systematic analysis of the ability of LLMs to generate and repair invariants; Pei et al.~\cite{canllm} enhance generation capabilities through fine-tuning; Chakraborty et al.~\cite{rankllm} introduce a re-ranking approach and can distinguish between correct and incorrect inductive invariants; Wu et al.~\cite{lam4inv} use bounded model checking to filter valid predicates from complete invariants generated by LLMs; Cao et al.~\cite{clause2inv} leverage LLMs only to generate literals, and use a counterexample-driven method to combine a complete loop invariant. LimICE is orthogonal to a wide range of these methods. For example, it can be combined with the re-ranking approach proposed by Chakraborty et al.~\cite{rankllm} to accelerate lemma search, or integrates the fine-tuning technology of Pei et al.~\cite{canllm} to enhance the literal-generation capabilities of LLMs.

\section{Conclusion}\label{sec: conclusion}

In this paper, we propose a new framework, Incremental ICE, for incremental synthesis using learning methods. Under this framework, we instantiate an efficient loop invariant synthesis tool called LimICE, which integrates LLMs with the ICE-DT method and can provide more structured guidance to LLMs. Meanwhile, we design a warm start mechanism as a complement to LLMs. Evaluations on 367 linear and 50 nonlinear benchmarks demonstrate that LimICE performs better than other state-of-the-art methods. Our work shows the advantages of the incremental approach over the monolithic approach from multiple perspectives.

\bibliographystyle{ACM-Reference-Format}
\bibliography{sample-base}

%%
%% If your work has an appendix, this is the place to put it.
% \appendix
% \section{Benchmark Details}\label{sec: Benchmark Details}

\end{document}